\documentstyle[preprint,aps]{revtex}

\input epsf

\begin{document}
\draft
\preprint{}

\def\d{\Delta}
\def\bpsi{\bar\psi}
\def\to{\tilde\omega}

\vskip-2truecm\title{Inclusions induced phase separation in mixed lipid film}

\author{P. Sens\footnote{present address: Institut Charles Sadron, 6 rue
Boussingault, 67083 Strasbourg, France\\ email: sens@ics.u-strasbg.fr}
and S. Safran}
\address{Department of Materials and Interfaces,\\ Weizmann Institute of
Science,
Rehovot 76100, Israel}
\maketitle
\begin{abstract}

\vskip-1truecm
The effect of rigid inclusions on the phase behavior of a film containing a
mixture of lipid molecules is investigated. In the proposed model,
the inclusion-induced
deformation of the film, and the resulting energy cost are strongly dependent upon the
spontaneous curvature of the mixed film. The spontaneous curvature is in turn strongly
influenced by the composition of film. This coupling between the film composition and the
energy per inclusion leads to a lateral modulation of the composition, which follows the
local curvature of the membrane. In particular, it is shown that the inclusion may induce
a global phase separation in a film which would otherwise be homogeneously mixed. The
mixed film is then composed of patches of different average composition, separated by the
inclusions. This process may be of relevance to explain some aspects of lipid-protein
association in biological membranes.

\end{abstract}


\narrowtext

\section{Introduction}

Our understanding of lamellar systems for which the basic element is a lipid bilayer has
greatly progressed in recent years (Nelson {\em et al}., 1989; Safran, 1994;
Sackmann and Lipowsky, 1995). Attempts to apply this knowledge to biological systems have
lead to many exciting and fundamental questions, such as the morphology of vesicles,
diffusion and transfer across membranes, or fusion and adhesion of membranes (Sackmann
and Lipowsky, 1995). Among the systems of interest in biology are membranes containing
proteins or other embedded molecules; the presence of such inclusions within the assembly
leads to interesting new phenomena. Such complexes are of great importance in many areas
of cell biology, since membrane proteins perform such fundamental functions as the
formation of pores in the membrane, the transfer of signals through the membrane, or the
binding of other proteins to the membrane (Alberts {\em et al}., 1994). These proteins can be
polymer-like coils tethered to the cell membrane (peripheral proteins), or more rigid
molecules embedded in the membrane (integral proteins). In the latter case, they are
often mostly hydrophobic, and shield themselves from water contact when buried within the
hydrophobic part of the lipid bilayer (see Fig.~1). In both cases, membrane proteins are
known to locally disrupt the order in the fluid membrane. 
Following the pioneering work of Huang (Huang, 1986), a large body of work has
been done on the physics of lipid film with inclusions (for a recent review, see
Goulian, 1996).
Several possible effects have been considered theoretically in order to
describe the behavior of the membrane protein complexes, such as a change of the membrane
rigidity and a tilt of the membrane at the location of the inclusion (Goulian, 1993), or
a local expansion of the bilayer to accommodate the inclusion size (Mouritsen and Bloom,
1984-1993; Dan, 1993-1994). The latter seems of particular relevance for hydrophobic
integral inclusions, and predicts that the deformation of the film around an inclusion
extends over a length related to the molecular details of the membrane, and leads to
non-monotonic interactions between neighboring inclusions.
In addition, several works have highlighted the influence of inclusions (e.g peptides)
on the phase behaviour of lipid solutions (Morein {\em et al}., 1997, May and Ben-Shaul,
1998).

Biological membranes are often formed by a wide variety of lipid and surfactant
molecules (Alberts {\em et al}., 1994). These mixed membranes have excited the imagination of
many theorists over the past years, because a coupling between the composition and the
shape of self assembled membranes promises many interesting phase transition phenomena
(Leibler, 1986; Leibler and Andelman, 1987; Seul and Andelman, 1995; Safran et al., 1991,
MacKintosh and Safran, 1993; Harden and MacKinstosh, 1994; Tanigushi, 1996; Gelbart and
Bruinsma, 1997). Most of those works exploit the coupling between film composition and
curvature related characteristics of membranes to connect a lateral phase separation and
a shape deformation of the membrane.

Despite extensive research, the role played by these different amphiphile species in real
(biological) systems is yet unclear.  Some evidence show that the composition of
amphiphile of a given portion of a cell membrane is related to the membrane proteins
present in the area (Alberts {\em et al}., 1994), a particular lipid environment being
needed in order for certain proteins to function properly. Explanations
invoking the lateral pressure exerted on a membrane protein by the lipid bilayer, as a
regulatory phenomenon for its morphology has been proposed to explain this specificity
(de Kruijff, 1997; Cantor, 1997a; Cantor, 1997b; Dan, 1998). On the other hand, the
existence of lipid domains in biomembranes is of fundamental importance to understand
specific membrane structures and functions. Domains of different compositions, which are
in the fluid state under physiological condition, are expected to be formed  by the
interplay of interactions between lipids and membrane proteins, and the mixing properties
of the lipids themselves (Melchior, 1986). 
These considerations have led to extensive experimental and theoretical research on the
behavior of mixed film with proteins. Some work follow the concept
of ``hydrophobic matching'', and bear some similarities with the present work
(Mouritsen and Bloom, 1993; Dumas {\em et al}., 1997 and reference therein). 
Several experimental studies have been performed to determined the influence of given
proteins on the stability of mixed membranes (Mittler-Neher and Knoll, 1989), but they
tend to focus mostly on electrostatic effects. Artificial mixed membrane are also useful
for the study of specific peripheral and integral membrane proteins close to
physiological conditions (Sanders and Landis, 1995). 

In this work, we investigate theoretically the consequences of the presence of inclusions
on the mixing properties of a mixed lipid film, assuming that an inclusion imposes a given deformation to the membrane. The aim of this paper is to extend previous results for the deformation and interaction
of fluid membranes with inclusions to bilayers containing two different species of lipid.
Special emphasis is given to the occurrence of phase separation in a mixed membrane which
would be homogeneously mixed in the absence of inclusion. The main results include the
prediction that the local composition of the film around an inclusion is controlled by the
deformation imposed on the film by the inclusion, and can vary widely. In our model, the
control parameter is the difference between the thickness of the film and the height of
the inclusion, but more complex couplings (slope, asymmetry..) could also be relevant,
and could be treated following the same pathway.

As in the often used ``Mattress model'' (Mouritsen and Bloom, 1984-1993), the lipid
molecules are modeled as spring with a given elasticity. In addition, an key ingredient
of the present model is the concept of curvature energy developed by Helfrich (Helfrich,
1973), and of particular importance is the idea of spontaneous curvature of a lipid
monolayer (Safran, 1994). It has been clearly demonstrated that the composition of a
mixed monolayer of amphiphilic molecules influences strongly its mean spontaneous
curvature (Milner {\em et al.}, 1989). We use this coupling between
composition and deformation of the membrane to derive the effect of a protein, not only
on the local composition near the inclusion, but also on the average composition of a
given patch of membrane. The two competing characteristics of the film, namely the
stretching and curvature of the lipid molecules, define a characteristic length over
which a membrane deformation vanishes. It defines also the range of the lipid
mediated interactions between inclusions, which is of order $100\AA$ in typical
biological membranes. It should be noted that in our model, the membrane is always
supposed to be in the fluid state. For this reason, this characteristic length
is a constant and one does not observe the temperature dependence which has been predicted
in relation to the gel-to-fluid transition in the membrane (Sperotto and Mouritsen,
1991). One advantage of our model over the mattress model is that our findings are
expressed with physical quantities (bending rigidity, spontaneous curvature) that could
be measure via independent experiments. Moreover, any more ``exotic'' deformations
imposed by the inclusion (such as a tilt of the bilayer interface), or direct, specific
interaction between the inclusion and one or the two types of lipid, can be introduced by
a straightforward change of the boundary conditions. Finally, this model predicts
a deformation of the membrane even if the inclusion hydrophobic core matches exactly the
thickness of the film, providing that the lipid monolayer has a non zero spontaneous
curvature.

The energy per inclusion being strongly influenced by the value of
the average spontaneous curvature of the film (it even change sign upon changing the
value of this parameter), it is argued that a ``global'' phase separation in the film may
be induced by the presence of inclusions for a certain range of parameters. Our results
show that some incompatibility between the two types of lipid is still needed to observe a
phase separation in the membrane, but that the presence of inclusions may increase the
critical temperature below which the membrane ceases to be an homogeneous 2-D
fluid. The film is then divided into patches of very different average compositions,
hence different mean spontaneous curvature, which create a nonrandom lipid distribution
in the bilayer.

Most molecular models describe the equilibrium conformation and energy of a lipid bilayer
as the interplay of three terms. Two surfaces terms, account for the interactions between
the hydrophilic heads of the molecules at the water interface, and the surface tension of
the film; the third, is a spring-like term for the stretching of the hydrophobic lipid
chains due to packing constraints (Ben-Shaul and Gelbart, 1995). The minimization of this
energy defines an optimum area per lipid head in the film, from which the equilibrium
bending and stretching characteristics of the film can be derived. In the following, we
will consider a two-dimensional symmetrical bilayer, and we focus only on the deformation
of one monolayer, assuming that the inclusion induce  deformations that are symmetric in
each monolayer. The deformation energy per unit length can be written in the general
form (Helfrich, 1973) (see Fig.~1 for notations):
\begin{eqnarray}
f=f_0+{1\over 2} k (u-&u&_\infty)^2+{1\over 2}\kappa u''^2 -\kappa
c_0 u''\nonumber\\
&u&'\equiv {du\over dz}
\label{en0}
\end{eqnarray}
where
$u$ is the deviation of the membrane surface from the flat state,
$k$ and $\kappa$ are the stretching and bending modulii of one monolayer per unit length,
and $c_0$ is the spontaneous curvature of the monolayer. Note that according to this
formula, a positive spontaneous curvature tends to give a concave bend towards the
outside of the monolayer. The reference energy, $f_0$, is the energy of a flat film. To
within a constant, this energy can be written (Safran, 1994) $f_0=(1/2) \kappa
c_0^2$. It reflects the fact that the monolayer is forced to be flat for symmetry
reasons, while it would rather adopt the curvature $c_0$.  For a mixed lipid film, the
three parameters
$k,\ \kappa$ and $c_0$ are functions of the local composition of the film. Rigorously
speaking, these parameters also depend on the local area per lipid head, which is, in
turn, coupled to the deformation. However, it will be shown in Section II that one need
not consider such a coupling to reach a correct physical picture of the effect of the
membrane inclusion.

In the next section, we will give a simplified presentation of previous results for a
one-component lipid film. We then study mixed lipid films containing by two types of
molecules of different spontaneous curvature. In section III, we derive the expression
for the deformation of the film interfaces, the local composition of the film, and the
interaction energy between two fixed inclusions. We study both the case of random mixing
and the case where some specific interactions exist between the two types of lipid. In
section IV, we discuss the possibility of a phase separation in mixed film. We show that
the presence of inclusion always favors the demixing of the two components. Nevertheless,
this effect alone is never strong enough to overcome the strong tendency for mixing which
is a property of the reference energy of a flat film without inclusions. However, if
there already exists a weak repulsive interaction between the two lipid species, the
inclusion induced tendency to phase separation may result in such a separation in a film
which would be homogeneous in the absence of inclusions

\section{Inclusion in a pure lipid bilayer}

In this section, we review in a simplified way the results in Ref. (Dan, 1993) for a
one component fluid bilayer with inclusions. The deformation of one monolayer interface,
and the resulting interaction energy between two immobile inclusions is calculated.  For
our purposes,  we neglect the coupling between the deformation and the area per lipid
molecule (Dan, 1993).

We assume that the only physical effect of the inclusion is to impose a given height of
the interface at contact, reflecting the fact that the inclusion is hydrophobic (see
Fig.~1). The energy per unit length Eq.(\ref{en0}) is normalized using the displacement
$\d=(u-u_\infty)/(u_0-u_\infty)$, where
$u_0$ is the height of the film at the inclusion\footnote{The case $u_0=u_\infty$ is also
of interest, since the inclusion allows for a discontinuity of the curvature of
the membrane interfaces. This case can be studied straightforwardly from our results} and
$u_\infty$ is the equilibrium thickness of the bilayer.   The total energy of a film
between two inclusions separated by
$2L$ is:
\begin{equation}
\Delta F\equiv F-F_0={k_BT L\over a}\Gamma\int_0^1 dx\left({\d''^2\over
\to^4}+\d^2-2
C_0{\d''\over\to^2}\right)
\label{en1}
\end{equation} 
where $a$ is a molecular size, and
$\Gamma={1\over2}k(u_0-u_\infty)^2a/(k_BT)$ measures is the ratio between the typical
stretching energy of the lipid tails and the thermal energy ($k_BT$). The normalized
coordinate is $x=z/L$, and $\to=\omega L$ with $\omega=(k/\kappa)^{1/4}$ being the inverse
of the characteristic length of the film, $C_0=c_0/((u_0-u_\infty)\omega^2)$ is the
normalized spontaneous curvature. The energy of the flat film becomes in these units:
$F_0=TL/a~\Gamma C_0^2$ (note that in these units, $\to$ varies linearly with the
separation between the two inclusions). The functional minimization of this energy  gives
a differential equation for the equilibrium profile of the film interface:
\begin{equation}
\d^{({\rm iv})}+\to^4\d=0
\label{bc0a}
\end{equation}

Two of the boundary conditions are physically fixed by the inclusions:
$\d|_{x=0}=1$ (to match the height of the inclusion) and $\d'|_{x=1}=0$ (by symmetry;
$x=1$ is at the midpoint between the of two inclusions). Two other boundary conditions
are set by minimization requirements, and read (Fox, 1950): $(\partial
f/\partial\d'')|_{x=0}=0$ and
$\partial_x(\partial f/\partial\d'')|_{x=1}=0$, where $f(\d,\d'')$ is the energy density.
Finally, the four boundary conditions are:
\begin{equation}
{\rm for}\ x=0\quad\cases{\d=1\cr
\d''=\to^2C_0\cr}
\qquad{\rm for}\ x=1\quad\cases{\d'=0\cr
\d'''=0\cr}
\label{bc0b}
\end{equation}

After proper integration by part and the use of the equations (\ref{bc0b}), the energy
per inclusion Eq.(\ref{en1}) can be written as a function of derivatives of
$\d$ at $x=0$ only:
\begin{equation}
\Delta F={k_BTL\over a}\Gamma\left({C_0\over \to^2}\d'(0)+{1\over
\to^4}\d'''(0)\right)
\label{ensimple}
\end{equation}

The deformation profile is easily derived:
\begin{equation}
\d(x)={(1-iC_0)\over 2}{\cosh{\left(\sqrt
i\to(1-x)\right)}\over\cosh{\left(\sqrt
i\to\right)}}+{(1+iC_0)\over 2}{\cosh{\left(i\sqrt
i\to(1-x)\right)}\over\cosh{\left(i\sqrt i\to\right)}}
\label{del0}
\end{equation} 
and the interaction energy between two inclusions separated by a distance
$2L$ is:
\begin{equation}
\Delta F=-{k_BTL\over a}\Gamma{\left(C_0^2-2 C_0-1\right)\sin{\sqrt2\to}+
\left(C_0^2+2 C_0-1\right)\sinh{\sqrt2\to}\over
\sqrt2\to\left(\cos{\sqrt2\to}+\cosh{\sqrt2\to}\right)}
\label{entot0}
\end{equation}

An estimate of the energy of interaction gives the value $\Delta F\sim 5 k_BT(\Delta
u/h)^2$ in the case of an hydrophobic mismatch $\Delta u$ between the inclusion and a
bilayer of thickness $2h$ (no spontaneous curvature), and the value $\Delta F\sim 5
k_BT(c_0h)^2$ for monolayers of spontaneous curvature $c_0$ and no mismatch. Both
energies should be higher than $k_BT$ and hence, represent relevant contributions to
aggregation/dispersion of inclusion in membrane, in addition to any direct physical
forces\footnote{For comparison, the magnitude of the Van der Waals attraction between
two cylinders of radius $R$ and separation $L$ in a medium of Hamaker constant $A\sim
20k_BT$ is $F\sim A/6 h R^{1/2}/L^{3/2}\sim k_BT$ (Israelachvili, 1992)}.

The profile of the film interface and this energy are shown in Figs.~2\&3 for typical values
of the parameters. The perturbation decreases over a length
$\to^{-1}$, which is also the range of the interaction between inclusions. It is a
measure of the relative strength of the stretching and the bending modulii. If the
stretching dominates, $\to$ is large and the profile decays rapidly to its asymptotic
values, and conversely. The spontaneous curvature fixes the curvature at
$x=0$ (the inclusion is assumed to constraint only the height of the film at that point),
and has a strong influence on the sign of the interaction energy, which can change upon
changing $C_0$ (see Fig.~3). These results are very similar to the conclusions of
Dan {\em et al}. (Dan, 1993), which tends to suggest that a complete physical picture of the
system can be obtained without considering the coupling between the film profile and the
area per lipid molecule.

As will be  shown in the next section, the phase behavior of a mixed lipid film is
closely related to the variation of the energy of an homogeneous film with the
spontaneous curvature $C_0$. From the expression of the energy Eq.(\ref{entot0}), one can
easily show that the energy of the film is non-monotonous, and reaches a maximum value at
a given value of
$\to$ for a spontaneous curvature:
\begin{equation}
\bar C_0={{\bar
c_0}\over(u_0-u_\infty)\omega^2}=-{\sinh{\sqrt2\to}-\sin{\sqrt2\to}\over
\sinh{\sqrt2\to}+\sin{\sqrt2\to}}
\label{cbar}
\end{equation}

\section{Inclusion in a mixed lipid bilayer}

We consider a mixed film of average composition $\bpsi$ of molecules $(A)$ and
$1-\bpsi$ of molecules $(B)$. For simplicity, we consider the case where these two types
of molecule differ only in their spontaneous curvature $c_A$ and
$c_B$ (their stretching and bending modulii being identical) and, as earlier, we need not
consider any change in the area per lipid head\footnote{Practically, parameters such as
the stretching and bending modulli are mostly influences by the hydrophobic tails of the
lipids, while the spontaneous curvature results of the asymmetry between heads and
tails (Israelachvili, 1992)}. We consider the free energy of the lipid film as a
function of two order parameters: the displacement of the film interface
$\d(x)$, and the variation of the film composition
$\phi\equiv\psi(x)-\bpsi$. The free energy can be written as a general Landau-Ginzburg
expansion up to second order in these order parameters\footnote{We have omitted quadratic terms without clear physical
meanings such as $\Delta''\phi'$.}, but all the terms
can be motivated from microscopic considerations as described in the Appendix:
\begin{equation}
\Delta F={k_BTL\over a}\Gamma\int_0^1dx\left\{{\d''^2\over
\to^4}+\d^2-2C_0{\d''\over\to^2}-
2{C_1\over\to^2}\left(\d''+\mu\right)\phi+D\phi^2+E
\phi'^2\right\}
\label{enbig1}
\end{equation} 
where the constants $D$ and $E$ are  positive if there is no
spontaneous phase separation in the film in the absence of foreign inclusions.

Such a coupling between the membrane deformation and local parameters of the film has
already been formulated in previous works, in particular to describe curvature
instability related to the presence of inhomogeneities in the membrane
(Leibler, 1986). The physical significance of the two first parameters $\to$ and
$C_0$ has been given in the previous section. The three new parameters
$C_1$, $D$, and $E$ can be explained physically with the aid of a molecular model for the
bilayer (see Appendix). The parameter $C_1$ gives the coupling between the spontaneous
curvature of the film and the local composition, and corresponds roughly to the
difference of spontaneous curvature between the two kinds of molecule. The parameter $D$
includes the mixing entropy of the film, specific interactions between lipids, and
spontaneous curvature terms, and the gradient term
$E$ arises from interactions between the different lipids; if the free energy of mixing
includes only entropic contributions, $E=0$. The chemical potential
$\mu$ is determined by the conservation relation:$\int_0^1\phi dx=0$.

In what follows, we first study the behavior of the film if the term in gradient of the
composition  vanishes in Eq.(\ref{enbig1}). Although not realistic, this treatment gives
us some insight into the more complete case ($E\ne 0$), treated in the next subsection.

\subsection{Mixed bilayer without interaction between different species}

For simplicity, we first consider the case where there are no specific interactions
between the two species and the gradient term in Eq.(\ref{enbig1}) vanishes ($E=0$). The
problem is then greatly simplified since the composition of the film can be readily
expressed as a function of the deformation
$\d(x)$ by minimizing Eq.(\ref{enbig1}). Invoking the conservation relation to determine
the chemical potential, one obtains:

\begin{equation}
\mu[\d]=\d'(0)\hskip 2truecm\phi[\d]={C_1\over \to^2
D}\left(\d''(x)+\d'(0)\right)
\label{muphi}
\end{equation} 
where the symmetry requirement $\d'(1)=0$ has been used. The equilibrium free energy can
then be expressed as a function of the interface profile only:
\begin{equation}
\Delta F={TL\over
a}\Gamma\int_0^1dx\left\{{\d''^2\over\to^4}+\d^2-2C_0{\d''\over\to^2}-
{\epsilon\over\to^4}\left(\d''+\d'(0)\right)^2\right\}
\label{entot}
\end{equation} 
where the difference between the two lipid species appears only in the perturbation
parameter
\begin{equation}
\epsilon={C_1^2\over D}
\label{epsilon}
\end{equation}

The Euler-Lagrange equation and the boundary conditions for the profile are obtained by a
functional minimization of (\ref{entot}):
\begin{equation}
\d^{({\rm iv})}+{\to^4\over (1-\epsilon)}\d=0
\label{eqtot0a}
\end{equation} 
with the four boundary conditions obtained similarly to Eq.(\ref{bc0b}):
\begin{equation}
{\rm for}\ x=0\quad\cases{\d=1\cr
(1-\epsilon)\d''=\to^2
C_0+\epsilon\d'\cr}
\qquad{\rm for}\ x=1\quad\cases{\d'=0\cr
\d'''=0\cr}
\label{eqtot0b}
\end{equation}

One can see from this set of equation that one of the main effects of the modulation of
composition is to reduce the bending modulus of a monolayer, as was already observed in
(Leibler, 1986). Hence, the modulation leads to a change of the characteristic
length scale: $\to^*=\to/(1-\epsilon)^{1/4}$ and of the normalized spontaneous curvature:
$C_0^*= C_0/\sqrt{1-\epsilon}$. One also notices a term involving $\d'(0)$ in the
boundary conditions. As the perturbation parameter $\epsilon$ approaches
$1$, the effective bending rigidity of the film becomes small and the quadratic form of
the energy in $\d''$ (which assumes a small curvature of the membrane everywhere) ceases
to be valid. This phenomenon has been referred to as ``Curvature Instability" in
asymmetric lipid bilayers (Leibler, 1986). It has also been identified as the
onset of the spontaneous formation of vesicles in mixed lipid bilayers (Safran
{\em et al}., 1991; MacKintosh and Safran, 1993).

Eq.(\ref{eqtot0a}) coupled with the boundary conditions Eq.(\ref{eqtot0b}) at $x=1$
imposes solutions of the form
\begin{eqnarray}
\d(x)=\sum_{i=1}^2 A_i&{}& \cosh{q_i(1-x)}\nonumber\\
{\rm with}\quad q_1=\sqrt{i}{\to\over(1-\epsilon)^{1/4}}&{}&\qquad
q_2=i\sqrt{i}{\to\over(1-\epsilon)^{1/4}}
\label{sol11}
\end{eqnarray} 
and corresponds to an interaction energy:
\begin{equation}
\Delta F={k_BTL\over a}\Gamma\left({C_0\over\to^2}\Delta'(0)+{(1-\epsilon)\over
\to^4}\Delta'''(0)\right)
\label{ensimple2}
\end{equation} quite close to the one derived for a pure lipid film (Eq.(\ref{ensimple}))

The full expression of the energy with the proper value of $\Delta$ and its derivatives
is not given here. As expected, the energy per inclusions is always lower when
modulations of the composition of the film are allowed.

\subsection{Including interaction between different species}

If the interaction between the different types of lipids are taken into account, the
problem is complicated by the fact that the concentration $\phi$ and the deformation
$\d$ are determined by two coupled differential equations. The functional minimization of
the energy (\ref{enbig1}) with respect to the two order parameters gives
\begin{equation}
\d^{({\rm iv})}+\to^4\d-\to^2 C_1\phi''=0 \qquad E \phi''-D\phi+{C_1\over
\to^2}(\d''+\mu)=0
\label{eqbiga}
\end{equation}
The boundary conditions are obtained similarly to Eq.(\ref{bc0b}), augmented by the
conditions: $(\partial f/\partial\phi')|_{x=0,1}=0$ (Fox, 1950):
\begin{equation}
{\rm for}\ x=0\quad\cases{\d=1\cr
\d''=\to^2 C_0+\to^2 C_1\phi\cr
\phi'=0}
\qquad{\rm for}\ x=1\quad\cases{\d'=0\cr
\d'''=0\cr
\phi'=0}
\label{eqbigb}
\end{equation}    
The value of the chemical potential can be obtained by the integration of the second
equation of (\ref{eqbiga}), between $x=0$ and $x=1$:
$\mu=\d'(0)$.

With the help of the Euler-Lagrange equations and the boundary conditions, we can express
the energy in a more compact form depending only on the profile at the inclusion.
Remarkably, it has the same expression as for a one component film (Eq.(\ref{ensimple})):
\begin{equation}
\Delta F={k_BTL\over a}\Gamma\left({C_0\over \to^2}
\d'(0)+{1\over\to^4}\d'''(0)\right)
\label{enbigcomp}
\end{equation}

After the change of variable
\begin{equation}
\phi(x)\rightarrow\Phi(x)\equiv\to^2C_1\phi(x)-\epsilon\d'(0)
\label{chvar}
\end{equation}
 Eq.(\ref{eqbiga}, \ref{eqbigb}) become
\begin{equation}
\d^{({\rm iv})}+\to^4\d-\Phi''=0 \qquad \lambda^2\Phi''-\Phi+\epsilon\d''=0
\label{eqbig2a}
\end{equation} 
where $\lambda\equiv\sqrt{E/D}$ is a lengthscale related to the variation of the
composition $\phi$ (and $\epsilon={C_1^2\over D}$). The boundary conditions become
\begin{equation}
{\rm for}\ x=0\quad\cases{\d=1\cr
\d''=\to^2 C_0+\Phi+\epsilon\d'\cr
\Phi'=0}
\qquad{\rm for}\ x=1\quad\cases{\d'=0\cr
\d'''=0\cr
\Phi'=0}
\label{eqbig2b}
\end{equation}
The Eq.(\ref{eqbig2a}), coupled with the three boundary conditions, Eq.(\ref{eqbig2b}),
for $x=1$, imposes solutions of the form:
\begin{eqnarray}
\d(x)=\sum_{i=1}^3A_i\cosh{q_i(1-x)}\qquad\Phi(x)&=&\sum_{i=1}^3B_i\cosh{q_i(1-x)}
\nonumber\\ 
B_i={\textstyle{A_i}\over\textstyle{\to^2 C_1}}{\textstyle{\epsilon
q_i^2}\over\textstyle{1-\lambda^2q_i^2}}&{}&
\label{root}
\end{eqnarray} 
where the wave vectors $q_i$ are defined by the equations:
\begin{equation} q_i=\sqrt{Q_i}\quad {\rm with}
\quad -\lambda^2 Q_i^3+(1-\epsilon)Q_i^2-\lambda^2\to^4 Q_i+\to^4=0
\label{root2}
\end{equation}

The three values of $A_i$ are determined by the boundary conditions, Eq.(\ref{eqbig2b}),
at $x=0$. 
Fig.~2 shows typical deformation and lipid concentration, where it is easily seen
that one of the two species is ``attracted'' toward the inclusion. Such an example of
surface enrichment has already been discussed in the context of the
hydrophobic matching model (Mouritsen and Bloom, 1993), and is clearly expected in our
situation as well. We would like however to reiterate here that all the results are
determined by energy minimization and that there is no assumption of any particular
composition near the inclusions, nor any length scale at the which the composition
changes.
 From Eq.(\ref{enbigcomp}), the
energy per inclusion takes the form:
\begin{equation}
\Delta F=-{k_B TL\over a}\Gamma\sum_i \left({\to^2 C_0+q_i^2\over \to^4}\right)
q_iA_i \sinh{q_i}
\label{engen1}
\end{equation}

The energy is now a function of four variables, namely the two length scale:
$\to^{-1}$ and $\lambda$, related to the variation of the displacement and the
composition, the spontaneous curvature $C_0$, and the coupling between the spontaneous
curvature and the composition of the film $\epsilon$. The explicit expression of the
energy of the film is not given here, but it is clear from Eq.(\ref{enbig1}) that this
energy always lies between the energy of an homogeneous film ($E\rightarrow\infty$ or
$\phi=0$), and the energy of a film without interaction ($E=0$). It should be noted at
this point that according to the molecular model given in the appendix, the two
parameters $\epsilon$ and $\lambda$ are not independent. Fig.~3 shows the plot of the
energy per inclusion, taking this coupling into account.

The full expression for the energy will be used in the next section to described the
phase separation process.

\section{Inclusion-induced phase separation in a mixed lipid film}

In this section, we describe the demixing of the film into patches of different average
composition, as driven by the presence of inclusions. We still restrict ourselves to two
dimensional films, although the existence of a line tension between different phases in
real (three dimensional) films, which can be avoided by the presence of an inclusion
between two phases in a 2-D film, is usually  an important factor for the size and shape
of the different domains (see Fig.~1b). Since we are only interested in the influence of
the inclusions as initiators of the transition, we do not consider any line tension term
here. Such a term would also be present in the description of the phase separation in a
mixed film without inclusions (Seul and Andelman, 1995).

In the previous sections, three different energies per inclusion have been calculated,
with increasing accuracy and increasing complexity. The first energy, Eq.(\ref{entot0}),
is the energy of a simple, one component, lipid film. An approximation for the energy of
a mixed film can be obtained from this equation by expressing the spontaneous curvature
$C_0$ as a function of the average composition
$\bar\psi$ of the film. This will be done in the next section. Secondly, the energy of a
mixed film has been derived, Eq.(\ref{sol11},\ref{ensimple2}), allowing for local
variation of the composition, $\phi=\psi(x)-\bar\psi$, but neglecting the interactions
between different species. From this calculation, a perturbation parameter $\epsilon$ has
been defined (Eq.(\ref{epsilon})). Finally, the interactions between different components
have been taken into account to obtain the third energy, Eq.(\ref{root}-\ref{engen1}),
which can be studied only via numerical analysis.

It is shown with a simple argument in the first paragraph below that although the
inclusions always favor the demixing of the film, the strong tendency for mixing due to
the spontaneous curvature for the flat film always dominates the curvature effects.
Hence, no phase separation is possible without repulsive interaction between the two
different species. A full study of the phase separation when these interactions are taken
into account is given in the next paragraph.

\subsection{Phase separation in a mixed film - simplified view}

We first study for simplicity the case where neither modulation of composition, nor
interactions between different species, are taken into account. This means that the
composition of the film is assumed to be everywhere equal to its average value
$\bar\psi$. The total free energy of the film is the sum of two contributions: the
reference energy of a flat film, including mixing entropy and mean field interaction
(Appendix, Eq.(\ref{ene0gen})), and the deformation energy due to the inclusion
Eq.(\ref{entot0}). Since we do not consider any specific interaction between lipid, the
parameters $\Delta c$ and $J$ are set to zero in the reference energy, which becomes:
\begin{equation} 
F_0={k_B T L\over
a}\left(\bpsi\log{\bpsi}+(1-\bpsi)\log{(1-\bpsi)}+
\Gamma C_0(\psi)^2\right)
\label{enref}
\end{equation} 
with $C_0(\psi)=C_A\psi+C_B(1-\psi)$, and where $\Gamma$ is the ratio of the stretching
energy at the inclusion to the thermal energy (see Appendix).

One can see that this reference energy strongly favors mixing, since both the entropy
term and the spontaneous curvature term are convex functions of the film composition.

The total energy is expressed as a function of $\bpsi$ from Eq.(\ref{entot0}):
\begin{eqnarray}
F={k_BT L\over a}\biggl(\bpsi
\log{\bpsi}+(1-&\bpsi&)\log{(1-\bpsi)}+\Gamma
C_1^2\Bigl(1-f(\to)\Bigr)\bpsi^2\biggr)+\alpha\bpsi+\beta\nonumber\\
0<f(\to)&\equiv&{\textstyle{\sin{\sqrt{2}\to}+
\sinh{\sqrt{2}\to}}\over
\textstyle{\sqrt{2}\to\left(\cos{\sqrt{2}\to}+\cosh{\sqrt{2}\to}\right)}}<1
\label{entot01}
\end{eqnarray}
where $\alpha$ and $\beta$ are some constants. The difference between the two
lipid molecules is expressed by the variable $\Gamma C_1^2$, which becomes in natural
units ${1\over2}\displaystyle{\kappa a c_1^2}/k_BT$, where $c_1=c_A-c_B$. The constant and
linear terms in this expression are as usual thermodynamically irrelevant since they are
related to a reference energy and a chemical potential.

In this simple case, the non-entropic contribution of the energy is quadratic in the film
composition $\bpsi$. The quadratic term is composed of a positive part, which favor
mixing and comes from the reference energy: $1/2
\kappa c_0^2$, and the negative part (which does favor phase separation), due to the
presence of inclusions. One can see that this term is always positive. This means that
although the presence of the inclusion favor a phase separation, the reference energy of
the flat film prevents the demixing.

According to Fig.~3, the energy is only slightly modified by a modulation of the local
composition of the film. This is especially true  for small separation between the
inclusions (small $\to$), which is the only situation where the effect of the inclusion
could compete with the reference energy. We thus expect our simplified view to still
hold, at least qualitatively, when we consider the modulation of the film composition and
the interactions between different species.

\subsection{Phase separation in a mixed film - full treatment}

In this section, we consider the full energy of a mixed film, including variations of the
local composition of the film, and interactions between the two lipid species. For
simplicity, we assume in what follows that the interactions between the
$A$ and $B$ lipids do not affect the spontaneous curvature of a mixed film of a given
composition. This means that $\Delta c=0$ in Eq.(A.1), and the spontaneous curvature is
then composed of a constant term and a term linear in $\phi$:
$c_0(\psi)=c_0(\bpsi)+c_1\phi$ with $c_1=c_A-c_B$.

We have seen in Eq.(\ref{entot}) that the modulation of the film composition introduce
the perturbation parameter:
\begin{equation}
\epsilon={1\over 1+\textstyle{1\over 2 \Gamma
C_1^2}\left(\textstyle{1\over\psi(1-\psi)}-J\right)}
\label{epsilon2}
\end{equation}
where, as already stated, $J$ is the interaction parameter between the two different
lipids, $C_1=c_1/(\omega^2(u_0-u_\infty))$, and $\Gamma={1\over
2}k(u_0-u_\infty)^2a/k_BT$ is the ratio between typical stretching and thermal energies.
Although this parameter stems from a calculation where the interaction between lipids are
not taken into account, it is reasonable to assume that this perturbation parameter is
still pertinent to the more complex case where the full energy is considered
(Eqs.(\ref{root}-\ref{engen1})).

When trying to calculate the influence of the inclusions on the value of the critical
demixing point, we run into the following problem. When modulations of the film
composition are allowed, the so-called curvature instability develops as
$\epsilon\rightarrow1$ (see discussion following Eq.(\ref{eqtot0b})). This occurs for the
value $J_{cr}=1/(\psi(1-\psi))$, which corresponds exactly to the position of the
spinodal curve for a model two fluid mixture, and is probably close to the spinodal curve
in our system as well. To properly describe  the inclusion-induced demixing with the
quadratic expansion Eq.(\ref{enbig1}), we need to be far enough from the onset of this
instability. For that reason, we choose to show the evolution of the critical interaction
energy $J$ (the binodal curve) for a given average composition
$\bpsi$ sufficiently far from
$\bpsi=1/2$, in which case binodal and spinodal curves are clearly distinct, and the
critical interaction parameter correspond to a value of $\epsilon<1$.

As already mentioned, the problem of the failure of the quadratic theory for
$\epsilon\simeq 1$ is reminiscent of the curvature instability
(Leibler, 1986). However, while those authors considered bilayers of
fixed thickness, but allowed for an asymmetry between the two sides of the bilayer, we
consider here a symmetrical bilayer allowed to change its thickness near a inclusions. The
inclusion not only fixes locally the thickness of the bilayer, but also allow for a
discontinuity of the curvature, which is impossible in membranes with no inclusions. If
the bilayer is asymmetrical, an overall spontaneous curvature can exist, giving rise of
rippled phase (Leibler, 1986) or spontaneous vesicles formation (Safran {\em et al}., 1991).

The binodal curve for a composition dependent energy $F[\psi]$ is obtained from the
so-called common tangent construction, leading to the two 
combined equations (Safran, 1994):
\begin{equation} 
{F[\psi_1]-F[\psi_2]\over\psi_1-\psi_2}=\left.{\partial F\over
\partial\psi}\right|_{\psi_1}=\left.{\partial F\over
\partial\psi}\right|_{\psi_2}
\label{binodal1}
\end{equation} 
which give the critical  value of $J=J_b$ for which regions of given composition
$\psi_1$ are in chemical and mechanical equilibrium with regions of composition
$\psi_2$.

If we assume an energy of the form : $F=\psi
\log{\psi}+(1-\psi)\log{(1-\psi)} + (\alpha/2)\psi^2+\beta \psi+\gamma$, the binodal
curve is given by the equations:
\begin{equation}
\alpha=-2{\log{1-\psi_1\over\psi_1}\over 1-2\psi_1}\quad{\rm and}\quad
\psi_2=1-\psi_1
\label{binodal2}
\end{equation}

A simplified expression of the binodal curve of a mixed film with inclusions can be
obtained using the simple treatment of the previous section (Eq.(\ref{entot01})), and
adding to the reference energy Eq.(\ref{enref}), a mean-field interaction term ${1\over
2}J\bpsi(1-\bpsi)$. Although somewhat artificial, since Eq.(\ref{entot01}) results from a
no-interaction calculation, this treatment gives a good approximation of the critical
demixing interaction, as is shown in Fig.~4. The resulting expression or $J_b$ is:
\begin{equation} 
J_b=2{\log{1-\bpsi_1\over\bpsi_1}\over
1-2\bpsi_1}+2\Gamma C_1^2\Bigl(1-f(\to)\Bigr)
\label{condsimple}
\end{equation}
where $f(\to)$ is defined in Eq.(\ref{entot01}). The condition $J>J_{b}$ (corresponding
to phase separation) is shown on Fig.~4. It is satisfied only if the repulsion between
the two components is strong enough to overcome {\it i)} the entropy of mixing, and {\it
ii)} the increase of the spontaneous curvature energy if the flat membrane phase
separates. The effect of the inclusions is simply to reduce the second requirement by a
factor of $\left(1-f(\to)\right)$ (Eq.\ref{entot01}). One clearly sees that the inclusions
reduce the slope of the boundary line, and that the effect becomes stronger as the
inclusions get close to each other.

The next step is to identify the effect of the inclusions on the variation of the full
energy $F[\bpsi]$ (Eq.(\ref{engen1})). This expression is far too complex to be
manipulated analytically. It is however possible to construct the binodal curve
numerically using a procedure equivalent to solving the Eqs.(\ref{binodal1}): We use the
Gibbs potential $G[\psi]=F[\psi]-\mu\psi$ where
$\mu$ is the chemical potential of the solution. For a given composition
$\psi_1\ll1/2$ (we choose $\psi_1=0.2$) we find the interaction $J$ and the potential
$\mu$ for which {\it i)} $G[\psi_1]$ is a minimum, and {\it ii)} there exists a
composition $\psi_2\gg1/2$ (practically $\psi_2\simeq1-\psi_1$) for which
$G[\psi_2]$ is a minimum. The numerical results are shown in Fig.~4 for a given
 value of the separation between inclusions, and agree well with the approximation given
by Eq.(\ref{condsimple}). One notable difference is that the binodal curve now depends
weakly on the absolute value of the spontaneous curvature (see the dependence on $C_B$ of
Fig.~4), and not only on the difference $C_1=C_A-C_B$. This is of course due to the fact
that the energy Eq.(\ref{engen1}) is not quadratic in the composition of the film, but
involves more complex couplings between curvature energy, mixing entropy, and specific
interactions.

\section{Summary}

The deformation of a mixed lipid film caused by the presence of embedded inclusions has
been predicted in term of the difference of spontaneous curvature between the lipid
molecules present in the film. The local spontaneous curvature is related to the local
composition of the film, and varies in order to match the film deformation. This
additional degree of freedom reduces the energy per inclusion with respect to the energy
of an homogeneous film, but does not modify it qualitatively. It is shown that the
composition of the membrane around a given inclusion is controlled by the deformation of
the film. For instance, inclusions larger than the film thickness would tend to be
surrounded by lipids of positive spontaneous curvature (see Fig.~2) while smaller
inclusions, which would pinch the film would be surrounded by lipids of negative
spontaneous curvature. This phenomenon could be partly responsible for the lateral
segregation in biological membranes, where nonrandom lipid distributions around given
membrane proteins have been experimentally demonstrated.

A particularly interesting phenomena is the inclusion-induced phase separation in the
mixed lipid film. Phase separation in a mixed fluid usually takes place when the
repulsive interaction between the two components is strong enough to overcome the
entropic mixing. In a mixed liquid film, there exists an additional effect due to the
coupling between the spontaneous curvature and the composition of the film. At the
simplest level, this coupling always favors mixing. We have shown in this paper that the
presence of inclusions can alter significantly this phase behavior.

Consider an homogeneously mixed film composed of two different lipid species. If the
(repulsive) molecular interactions between different constituents are increased, or the
temperature of the sample lowered, a phase separation can occur where regions of
different lipid composition are observed. We have shown that this phase separation can
also be triggered by the deformation  and composition modulation induced by intermembrane
inclusions. As the density of inclusions in the film is increased, the distance between
inclusions decreases up to the point where it reaches a critical value where the phase
separation occurs. In previous works (Dumas {\em et al}. 1997), computer simulations
have led to the conclusion that inclusions will preferentially be located at the interface
between two coexisting phases. We claim here that the inclusions can actually be
responsible for the very existence of the two phases.

The influence of the inclusions on the phase separation is shown in Fig.~4. They always
favor the phase separation, but their effect is never strong enough to compete with the
tendency of mixing of a flat film in the absence of other repulsive interactions, of
strength $J$. Nevertheless, the presence of inclusions can lower significantly the
critical interaction energy $J$ for which the phase separation occurs if the inclusions
are close enough to each other (typically
$\omega L<5$ where $\omega$ is the inverse decay length of the deformation).

Our model is based on a unidimensional lipid membrane, which, as was already mentioned,
is formally equivalent to considering infinitely long inclusions. This may seem a
very restrictive assumption, however, an analogy to the (more realistic) situation where
inclusions are closer to small cylinders (with a circular basis) in a 2D membrane is quite
straightforward. As can be seen when one compares our results for a single component film
(section II) with those obtained by the resolution of the full 2D problem (Dan 1994), a
good picture of the 2D situation can be obtained by dividing the bending and stretching
coefficient by a molecular size equivalent to the square root of the area per lipid head,
and by multiplying the energies per inclusion by $2\pi$. This should be the case in our
mixed film as well.

To conclude this work, we estimate the effect of the inclusions on the critical
interaction parameter for typical lipids. According to Eq.(\ref{condsimple}), the
inclusions reduce the interaction energy by the amount $2\Gamma C_1^2 f(\to)$, which, in
natural units, becomes $\kappa a c_1^2/(k_BT)f(\omega L)$. One first remarks that the
effect does not depend on the mismatch between the height of the inclusion
and the thickness of the lipid bilayer. Even inclusions matching the thickness of the
bilayer have an effect, since they allow for a discontinuity of the curvature of the
layer (Dan, 1993). The first important parameter is the ``healing length'' for the
deformation of the bilayer $\omega^{-1}$. Because of the quarter power of the dependence
of $\omega$ on the stretching and bending modulii of the lipid layers (see
Eq.(\ref{en1})), this length is only weakly dependent of the characteristics of the
membrane: $\omega^{-1}=0.5-2.5~h$ where $h$ is the thickness of one lipid monolayer $\sim
25{\rm\AA}$. The composition of lipid and protein molecules varies widely in biological
membranes. Typical protein over lipid ratios are of the order of $1/50$ (Alberts {\em et
al}., 1994), in which case the mean separation between proteins can be of order
$50-100{\rm\AA}$ (the area per lipid head is typically
$a^2=40{\rm\AA}^2$). Hence, in these membranes with high protein concentration, we can
reach values of
$f(\omega L)\sim 1/2$. The spontaneous curvature of a monolayer is typically the inverse
of its thickness: $c_1\sim h^{-1}$, and its bending rigidity of order
$\kappa\sim 25h~k_BT$ (in our 2-D model). Hence, the presence of proteins in a lipid
bilayer under biological conditions can reduce the critical interaction parameter
$J_b$ by an amount $1.5~k_BT$, which corresponds to $30\%$ of its value without
inclusions. We can thus expect many situations where the proteins not only control the
local composition of the membrane, but are also able to induce a major demixing of the
lipid membrane.

We are grateful to N. Dan and U. Alon for useful discussions. We wish to acknowledge the
donor of the Petroleum Research Fund administered by the American Chemical society, the
Israel Science Foundation, and the U.S. - Israel Binational Science Foundation for
partial support of this research.

\appendix

\section{About the Coefficients in the L-G expansion}

The coefficients in the Landau-Ginzburg expression, Eq.(\ref{enbig1}), can be related to
molecular parameters of the film. If one considers that the spontaneous curvature of a
monolayer arises principally from interactions between neighboring molecules, it can be
expressed as (Safran, 1994):
\begin{equation} 
c_0(\psi)=c_A\psi^2+c_B(1-\psi)^2+(c_A+c_B+\Delta
c)\psi(1-\psi)
\label{spont1}
\end{equation} 
where $\Delta c$ represents the contribution to the spontaneous curvature of neighboring
$A$ and $B$ molecules. For a weak modulation of the composition $\phi\ll\bpsi$, the
expression becomes:
\begin{eqnarray} 
c_0(\bpsi+\phi)=c_0(\bpsi)+c_1 \phi-\Delta c \phi^2\cr  {\rm
with}\quad c_1=c_A-c_B+\Delta c(1-2\bpsi)
\label{spont2}
\end{eqnarray} 
The expansion of the energy per unit length of a flat film gives
\begin{eqnarray} 
&{1\over 2}\kappa c_0(\bpsi+\phi)^2={1\over 2}\kappa
\left(c_0(\bpsi)^2+2c_0c_1\phi+c_2^2\phi^2\right)\cr  &c_2^2=c_1^2-2c_0\Delta c
\label{spont3}
\end{eqnarray}

The mixing energy per unit length, $f_m$, is written in the general form
(Safran, 1994):
\begin{eqnarray} 
f_m = {k_BT\over
a}\left(\psi\log{\psi}+(1-\psi)\log{(1-\psi)}+{1\over 2}J\psi(1-\psi)+{1\over
2}B|\nabla\psi|^2\right)\cr B={J\over2}\left({a\over L}\right)^2\hskip 5truecm
\label{mix1}
\end{eqnarray}  
where $J$ is the interaction parameter between the two species, and the gradient is a
derivative over the dimensionless variable $x=z/L$.

After expansion of all the term up to the quadratic order in the deformation and the
composition difference, the energy takes the form of Eq.(\ref{enbig1}), with
\begin{eqnarray} 
C_\alpha={c_\alpha\over(u_0-u_\infty)\omega^2}\qquad
D=C_2^2+\Gamma S_2\qquad E=\Gamma\left({a\over L}\right)^2{J\over 4}\cr
\Gamma\equiv{{1\over 2}k(u_0-u_\infty)^2 a\over k_BT}\quad
S_2\equiv{1\over2\bpsi(1-\bpsi)}-{J\over 2}
\label{def}
\end{eqnarray} 
The reference energy is:
\begin{equation} F_0={k_BT L\over
a}\left(\bpsi\log{\bpsi}+(1-\bpsi)\log{(1-\bpsi)}+{1\over
2}J\bpsi(1-\bpsi)+{C_0(\bpsi)^2\over\Gamma}\right)
\label{ene0gen}
\end{equation}

\begin {thebibliography}{99}
\end {thebibliography}
   \parindent=0pt \leftskip=0truecm
   \parskip=8pt plus 3pt \everypar{\hangindent=0pt}

Alberts, B., D. Bray, J. Lewis, M. Raff, K. Roberts, and J.D. Watson. 1994.
Molecular Biology of the Cell, Garland, New York.

Ben-Shaul, A. and Gelbart, W. 1994. Statistical Thermodynamics of Amphiphile
Self-Assembly: Structure and Phase Transition in Micellar Solutions. {\em In} Micelles,
Membranes, Microemulsions, and Monolayers, Gelbart, W.; Ben-Shaul, A.; Roux, D. editors.
Springer-Verlag, New York

Cantor, R. 1997a. Lateral Pressures in Cell Membranes: A Mechanism for Modulation of
Protein Function. {\it J. Phys. Chem. B}. 101:1723-1725

Cantor, R. 1997b. The lateral Pressure Profile in Membranes: A Physical Mechanism of
General Anesthesia. {\it Biochemistry}. 36:2339-2344 

Dan, N.; Pincus, P.; Safran, S. 1993. Membrane-Induced
Interactions Between Inclusions. {\em langmuir}. 9:2768-2771

Dan, N.; Bergmann, A.; Safran, S.; Pincus, P. 1994. Membrane-Induced
Interactions Between Inclusions. {\em J. Phys, II}. 4:1713-1725

Dan, N.; Safran, S. 1998.  Effect of Lipid Characteristics  on the Structure of
Trans-Membrane Proteins, {\it Biophysical Journal}. In press.

de Kruijff, B. 1997. Biomembranes - Lipids Beyond the Bilayer. {\em Nature}. 386:129-130

Dumas, F., Sperotto, M., Lebrun, M.-C., Tocanne, J.-F., Mouriten, O. 1997. Molecular
Sorting of Lipids by Bacteriorhodopsin in
Dilauroylphosphatidylcholine/ Distearoylphosphatidylcholine Lipid Bilayers. {\em Biophys.
J.} 73:1940-1953

Fox, C. 1950. An introduction to the Calculus of Variations. Oxford University
Press. Oxford

Gelbart, R.; Bruinsma, R. 1997. Compositional-Mechanical instability of Interacting
Mixed Lipid Membranes {\em Phys. Rev. E}. 55:831-835

Goulian, M.; Bruinsma, R.; Pincus, P. 1993, Long-Range Forces in
Heterogeneous Fluid Membranes. {\em EuroPhys. Lett.} 22:145-150, {\em ibid}.
Erratum. 23:55

Goulian, M. 1996. Inclusions in Membranes. {\em Current Opinion in Colloid \& Interfac
Science} 1,358-361

Harden, J.; MacKintosh, F. 1994. Shape transformations of Domains in Mixed-Fluid Films
and Bilayer Membranes. {\em EuroPhys. Let.} 28:495-500

Helfrich, W. 1973. Elastic Properties of Lipid Bilayers: Theory and Possible Experiments.
{\em Z. Natureforsch.} 28c:693-703

Huang, H. 1986. Deformation Free Energy of Bilayer Membrane and its Effect on Gramicidin
Channel Lifetime. {\em Biophys. J} 57:1075-1084

Israelachvili, J. 1992. Intermolecular \& Surface Forces, Chap.~17. Academic Press. San
Diego

Leibler, S. 1986. Curvature Instability in Membranes. {\em J. Physique}. 47:507-516

Leibler, S and Andelman, D. 1987. Ordered and Curved Mesostructures in Membranes and
Amphiphilic Films. {\em J. Physique}. 48:2013-2018

MacKintosh, F. and Safran, S. 1993. Phase Separation and Curvature of Bilayer Membranes.
{\em Phys. Rev. E}. 47:1180-1187

May, S. and Ben-Shaul, A. 1998. Molecular Theory of Lipi-Protein Interaction and the
$L_\alpha-H_{II}$ Transition. {\em Preprint}

Melchior, D. 1986. Lipid Domains in Fluid Membranes: A Quick-Freeze Differential
Scanning Calorimetry Study. {\it Science}. 234:1577-1580

Milner, S.; Witten, T. and Cates, M. 1989. Effects of POlydispersity in the End-Grafted
Polymer Brush. {\em Macromolecules} 22:853-861

Mittler-Neher, S.; Knoll,W. 1989. Phase Separation in Bimolecular Mixed Lipid Membranes
Induced by Polylysine. {\it Biochem. Biophys. Res. Commun.} 162:124-129

Morein, S. Strandberg, E., Killian. A., Persson, S. Ardvidson, G., Koeppe, R. and
Linblom, G. 1997. Influence of Membrane-Spanning $\alpha$-Helical Pptides on the Phase
Behavior of the Dioleoylphosphatidylcholine/Water System. {\em Biophys. J.} 73:3078-3088

Mouritsen, O. and Bloom, M. 1993, Models of ipid-Protein Interactions in Membranes. {\em
Annu. Rev. Biophys. Biomol. Struct.} 22:145-171

Mouritsen, O. and Bloom, M. 1984, Mattress Model of Lipid-Protein Interactions in
Membranes. {\em Biophys. J.} 46:141-153

Sperotto, M. and Mouritsen, O. 1991, Monte Carlo simulation studies of lipid order
parameter profiles near integral membrane proteins, {\em Biophys. J.}, 59:261-270

Nelson, D; Piran, T; Weinberg, S. editors.
1989. Statistical Mechanics of Membranes and Surfaces. World Scientific. Singapore.

Sackman, E. and Lipowky, R.; editors. 1995. Structure and Dynamic of Membranes. North
Holland. Amsterdam.

Safran, S.; Pincus, P.; Andelman, D.; MacKintosh, F. 1991. Stability and Phase Behavior
of Mixed Surfactant Vesicles. {\em Phys. Rev. A}. 43:1071-1077

Safran, S. 1994. Statistical Thermodynamics of Surfaces, Interfaces,
and Membranes. Frontiers in Physics. Addison-Wesley.

Sanders, C. and Landis, G.1995. Reconstitution of Membrane Proteins into Lipid-Rich
Bilayerd Mixed Micelles for NMR Studies. {\it Biochemistry.} 34:4030-4040

Seul, M. and Andelman, D. 1995. Domain Shape and Patterns; The Phenomenology of Modulated
Phases. {\em Science}. 267:476-483

Taniguchi, T. 1996. Shape Deformation and Phase Separation Dynamics of Two-Component
Vesicles. {\em Phys. Rev. Let.} 76:4444-4447

\newpage
\begin{figure}[h]
\vskip2truecm
\centerline{ \epsfxsize=10truecm \epsfbox{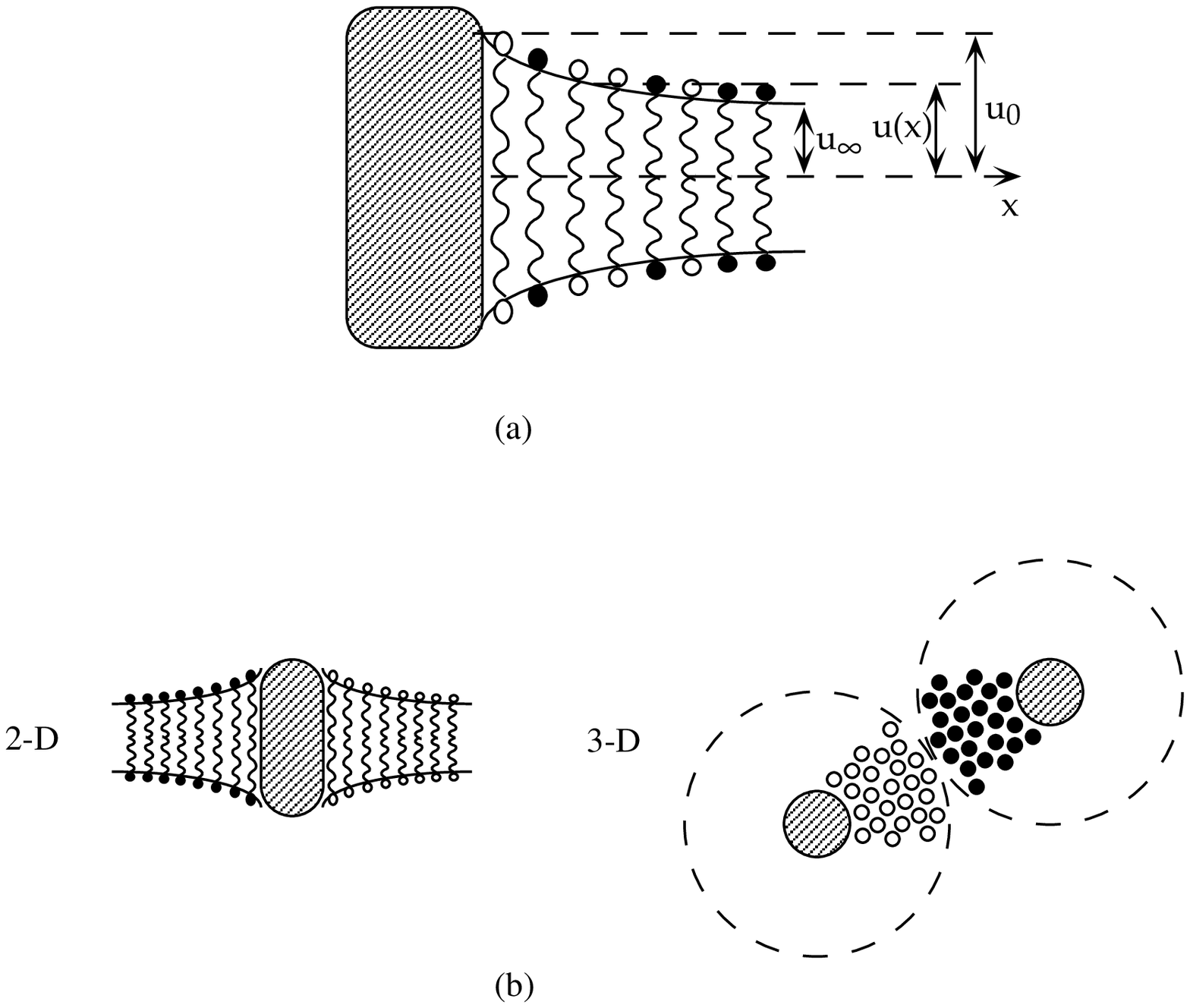} }
\vskip2truecm
\caption{\protect\small (a) Schematic representation of a mixed lipid film with inclusions. The
notations used in the text are described on the figure.  (b) Difference between the two
dimensional and three dimensional films for the phase separation process. In the 2D film,
phase separation can occur without the existence of a line tension, a situation that is
impossible in the 3D film. As noted in the next, the problem does not concern us as long
as we are  only interested in the effect of the inclusion on the phase separation.
\label{f1}}
\end{figure}

\newpage
\begin{figure}[h]
\centerline{\hskip0truecm\epsfxsize=17truecm \epsfbox{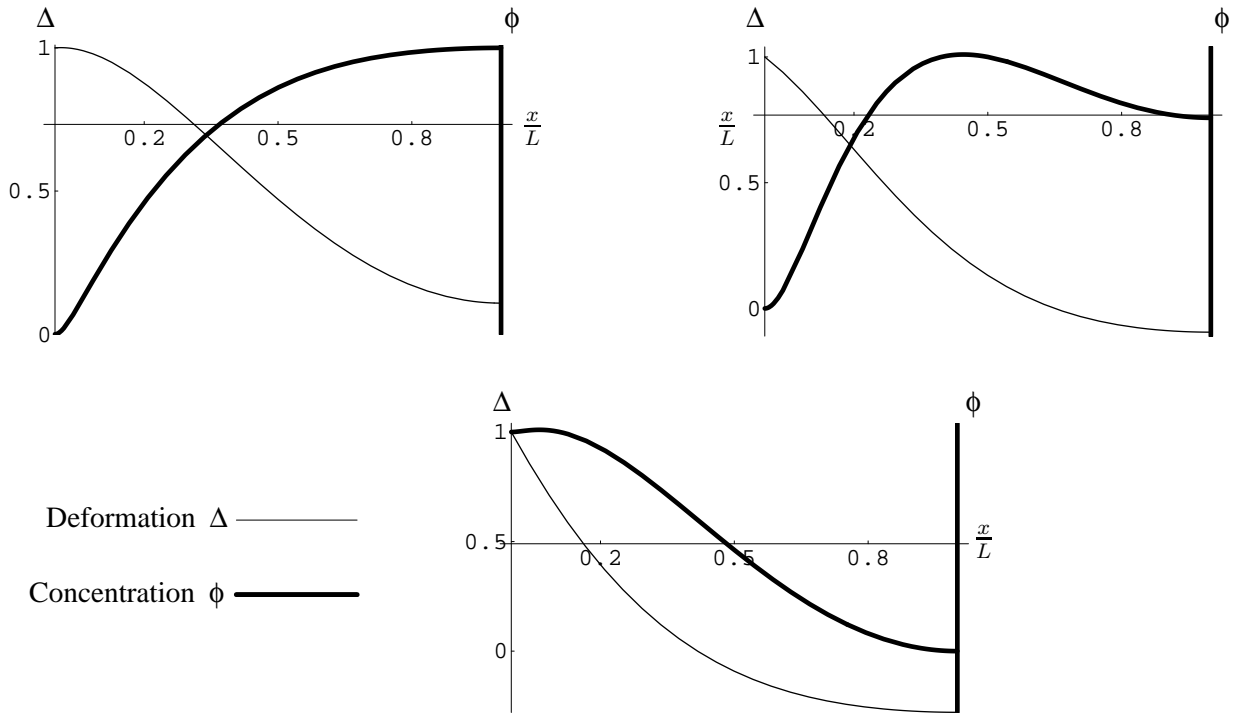}}
\vskip-9.3truecm\hskip7truecm${x\over L}$\hskip2.3truecm${x\over L}$
\vskip4.7truecm\hskip13truecm${x\over L}$
\vskip4.6truecm
\caption{\protect\small Effect of the inclusion on the deformation profile of a monolayer $\Delta(x)$ (thin curve, left axis) and on the composition of the film
$\phi(x)$ (thick curve, right axis), with the distance to the inclusion $x$, for three
different values of the spontaneous curvature $C_0=-0.8$ (upper left), $C_0=0$ (upper
right) and $C_0=0.8$ (lower). In this example, $C_1>0$, and the $A$ molecules move toward the
regions of positive curvature.
\label{f2}}
\end{figure}

\newpage
\begin{figure}[h]
\centerline{\epsfxsize=15truecm \epsfbox{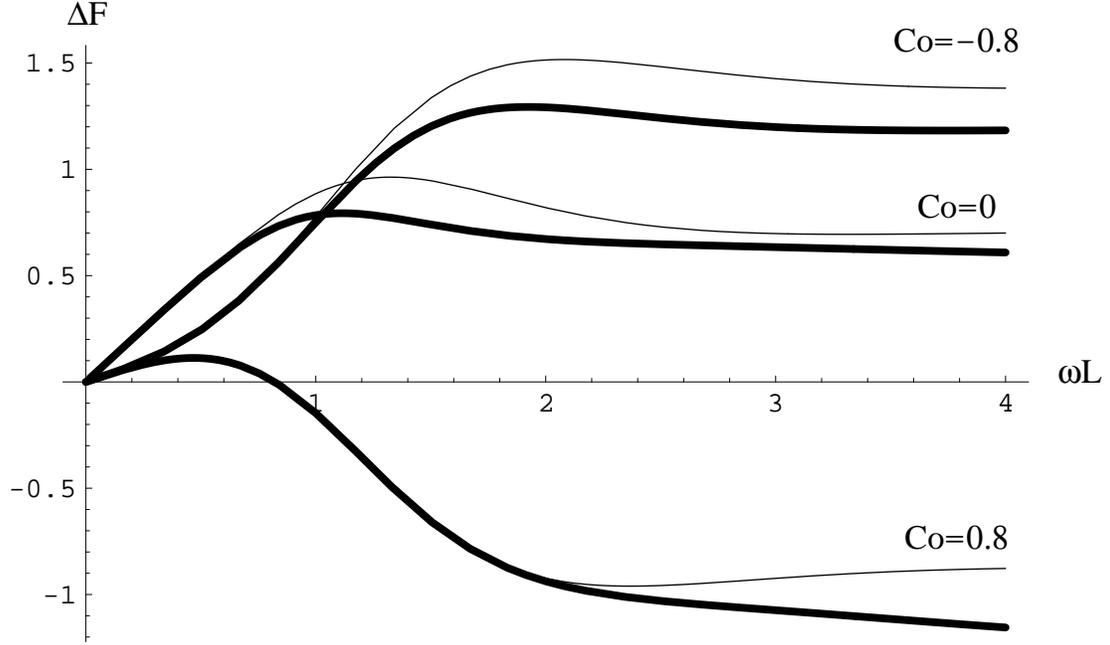}}
\vskip2truecm
\caption{\protect\small Variation of the interaction energy between two
inclusions with the separation between inclusions $\to=\omega L$ for the same three
values of the spontaneous curvature. The thin curves are plotted using the expression of the energy for a pure lipid film (Eq.7), and the fat curve is using the energy of a mixed
film (Eq.26), with $\Gamma=1$ and $a/L=1/20$. Note that the energy is attractive if
$C_0>0$.
\label{f3}}
\end{figure}

\newpage
\begin{figure}[h]
\centerline{\epsfxsize=18truecm \epsfbox{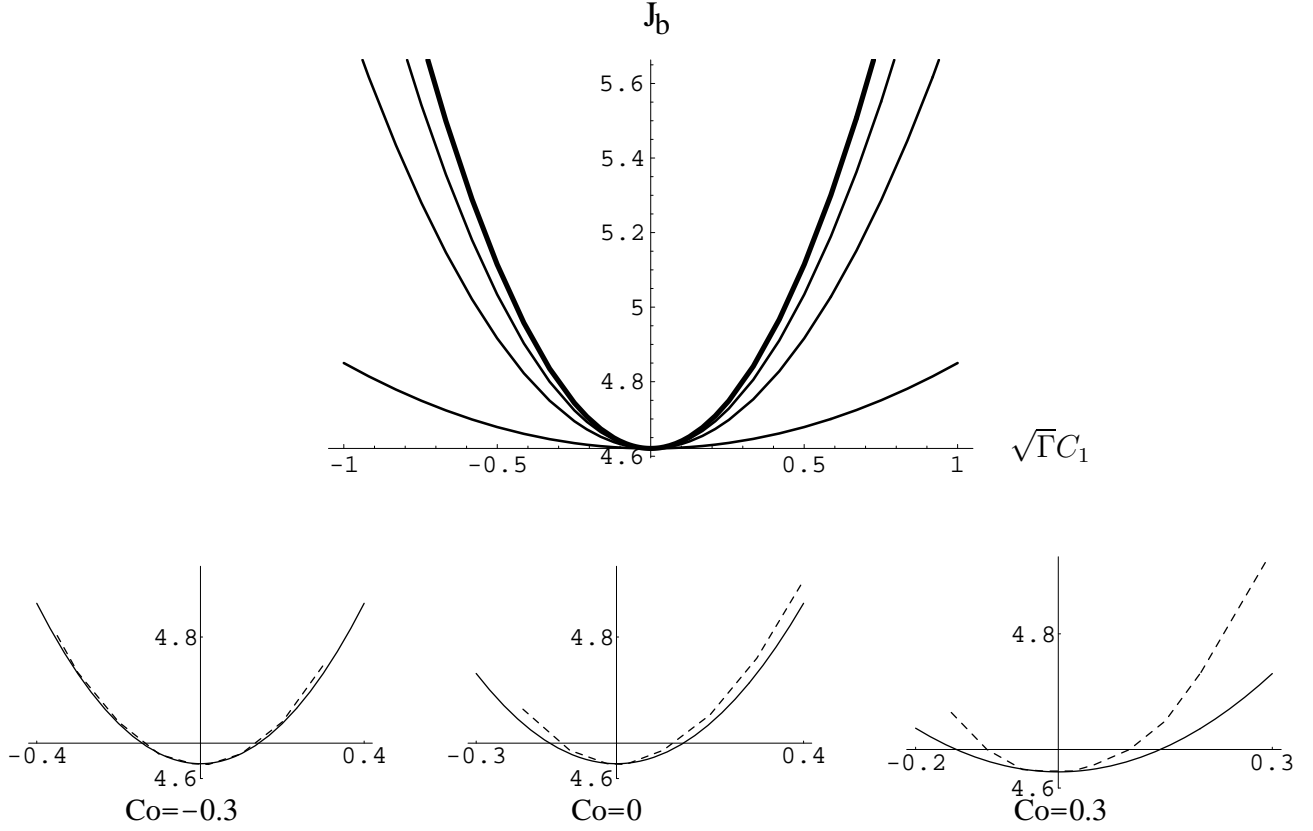}}
\vskip-5.8truecm\hskip13truecm$\sqrt{\Gamma} C_1$
\vskip7.8truecm
\caption{\protect\small Phase separation diagram. (a) Using the simplified expression Eq.(32). The
x-axis represent the spontaneous curvature difference between the two molecules: $C_1$ 
and the y-axis the repulsive interaction energy:
$J$. The thick line represents the boundary for a film without inclusion and the other
lines correspond to different separations between inclusions: $\to=1$,
$\to=2$ and $\to=4$. (b) Comparison between the previous approximation (full line) and the result
obtained numerically from Eq.(26) (dashed curves), for three values of the spontaneous
curvature of the $b$-species $C_B$. The quality of the fit depends of $C_B$..
\label{f4}}
\end{figure}

\end{document}